\documentstyle[aps,multicol,epsf]{revtex}
\begin{document}
\preprint{SNUTP }
\draft
\title{Facet Formation in the Negative Quenched Kardar-Parisi-Zhang Equation\\}

\author{H.~Jeong$^{\dag}$, B.~Kahng$^{\ddag}$, and D.~Kim$^{\dag}$}
\address{
     \dag Center for Theoretical Physics and Department of Physics, \\
          Seoul National University, Seoul 151-742, Korea \\
     \ddag Department of Physics and Center for Advanced Materials and Devices,\\
           Kon-Kuk University, Seoul 143-701, Korea \\}
\maketitle
\thispagestyle{empty}

\begin{abstract}
The quenched Kardar-Parisi-Zhang (QKPZ) equation with negative non-linear
term shows a first order pinning-depinning (PD) transition as the driving
force $F$ is varied. We study the substrate-tilt dependence of the dynamic 
transition properties in 1+1 dimensions. 
At the PD transition, the pinned surfaces form a facet
with a characteristic slope $s_c$ as long as the substrate-tilt $m$ 
is less than $s_c$.
When $m<s_c$, the transition is discontinuous and the critical value of the
driving force $F_c(m)$ is independent of $m$, while the transition 
 is continuous and $F_c(m)$ increases with $m$ when $m>s_c$. 
We explain these features from a pinning mechanism 
involving a localized pinning center
and the self-organized facet formation.
\end{abstract}


\begin{multicols}{2}
\narrowtext

The pinning-depinning (PD) transition 
by an external driving force has been of much interest recently. 
Typical examples are interface growth in porous media under external 
pressure \cite{porous}, dynamics of a domain wall under random fields
\cite{rfield,bausch}, dynamics of a charge density wave 
under an external field \cite{cdw}, 
and vortex motion in superconductors under external current 
\cite{vortex,ertas}. In the PD 
transition, there exists a critical value $F_c$ 
of the driving force $F$, such that
when $F < F_c$, interface (or charge, or vortex) is pinned 
by disorder, while for $F > F_c$, it moves with a constant 
velocity $v$. The velocity $v$ plays the role of order 
parameter in the PD transition, which typically behaves as 
\begin{equation} v \sim (F-F_c)^{\theta}.
\end{equation}

Recently, several stochastic models for the 
PD transition of interface growth in disordered media 
have been introduced \cite{boston,tang}.
It is believed that the models in 1+1 dimensions are described
by the quenched Kardar-Parisi-Zhang (QKPZ) equation for the surface height $h$,
\begin{equation}\label{qkpz}
{\partial_t h }=\nu{\partial_x^2 h } 
+{\lambda \over 2}( \partial_x h)^2 +F+\eta(x,h),   
\end{equation}   
where $\nu$ and $\lambda$ are constants and the noise $\eta$ 
depends on position $x$ and height $h$ with the properties of
$\langle \eta(x,h) \rangle =0$ and 
$\langle \eta(x,h)\eta(x',h') \rangle =2D \delta(x-x')\delta(h-h')$.
The QKPZ equation for $\lambda > 0$ exhibits a PD  
transition with $\theta \sim 0.64$. The surface at $F_c$ 
can be described by the directed percolation (DP) cluster 
spanned perpendicularly to the surface growth direction 
in 1+1 dimensions. The roughness exponent $\alpha$ of 
the interface is given as the ratio of the two correlation length exponents 
of the DP cluster, that is $\alpha=\nu_{\perp}/\nu_{\parallel}\approx 0.63$.
The coefficient $\lambda$ of the non-linear term renormalizes to $\lambda_r$
which scales as $\lambda_r \sim (F-F_c)^{-\phi}$ ($\phi \sim 0.64$). 
This is obtained
by measuring the substrate-tilt dependence of the velocity
\begin{equation}\label{vfm}
v(F,m) \sim v(F,0) + {\lambda_r \over 2} m^2
\end{equation}
where $m$ is the tilt of the initial substrate \cite{abs}.\\

Origin of the non-linear term in the QKPZ equation 
is different from that of the thermal KPZ equation with noise 
$\eta(x,t)$ \cite{dhar}. For the quenched case, the non-linear term 
is induced by the anisotropic nature of disordered media, 
while for the thermal case, it is induced 
by the lateral growth effect and is proportional 
to the velocity of the interface 
which vanishes at the threshold of the PD transition \cite{kpz}. 
In particular, Tang et al. \cite{dhar} showed in the context of
vortex dynamics that the effective pinning strength takes the form
$(\Delta_h+s^2\Delta_x)^{2/3}/(1+s^2)^{2/3}$ where $s=\partial_x h$
is the local slope and $\Delta_h^{1/2}$ and $\Delta_x^{1/2}$ are
the amplitudes of random forces in the $h$ and $x$ directions, respectively.
When the medium is anisotropic, $\Delta_h \ne \Delta_x$ in general and
the effective pinning strength depends on the local slope 
and generates the non-linear term in Eq.(\ref{qkpz}) with 
$\lambda \propto (\Delta_h - \Delta_x)$.
Therefore when $\Delta_h < \Delta_x$, i.e. when the surface is driven along
the easy direction, $\lambda$ is negative. \\

\begin{figure}
\centerline{\epsfxsize=7.3cm \epsfbox{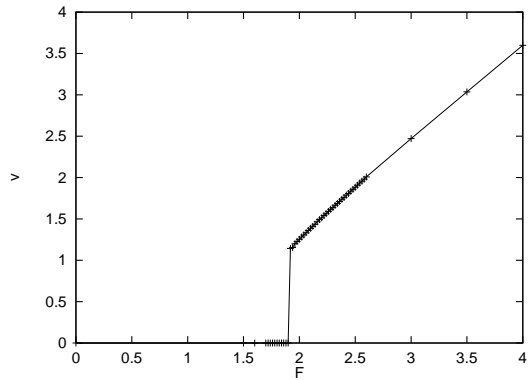}}
\caption{The velocity versus force to show the first order 
transition. The data are obtained from a flat substrate 
with $\lambda=-0.5$. The solid line is a guide to the eye.}
\label{fig1}
\end{figure}

\begin{figure}
\centerline{\epsfxsize=7.3cm \epsfbox{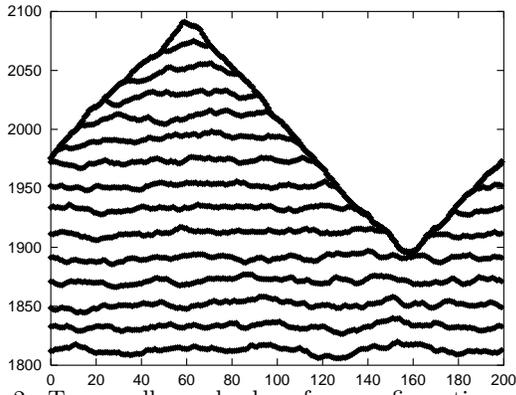}}
\caption{Temporally evolved surface configurations of the QKPZ 
equation with $\lambda=-0.5$ in 1+1 dimensions at the 
pinning-depinning transition point. Each curve is the surface profile
at constant time intervals. The facet slope is $s_c \approx 2.1$.}
\label{fig2}
\end{figure}

The negative QKPZ equation, i.e. the QKPZ equation with negative
$\lambda$ is first studied in \cite{noise}. In marked contrast to the 
$\lambda > 0$ case, the negative QKPZ equation describes a first order
transition in the sense that the velocity $v$ shows a discontinuous jump at
the PD transition. This is shown in Fig.\ref{fig1}. At the PD transition,
the pinned surface takes the shape of a mountain with flat inclination as
shown in Fig.\ref{fig2}. To elucidate the transition mechanism of the
negative QKPZ equation, we study in this work the substrate-tilt dependence
of the dynamic properties. 
We performed the direct numerical integration of Eq.(\ref{qkpz})
in one dimension with the discretized version used in \cite{noise}:
\begin{eqnarray}
h(x,&t&+\Delta t)=h(x,t)+\Delta t \{h(x-1,t)+h(x+1,t) \nonumber \\
&&-2h(x,t)+{\lambda \over 8}(h(x+1,t)-h(x-1,t))^2+F \} \nonumber \\
&&+(\Delta t)^{2/3}\xi(x,[h(x,t)]),
\end{eqnarray}
where [$\cdots$] denotes the integer part, and $\xi$ is
uniformly distributed in [${-{1 \over 2}}$, $1 \over 2$].
The prefactor $(\Delta t)^{2/3}$ of the noise term arises from approximately
coarse-graining the noise $\eta(x,h)$ during a time interval $\Delta t$.
This is different from that used in Ref.\cite{vicsek}, but
does not alter physical property of the PD transition. We choose 
$\Delta t=0.01$ and use 
the initial condition $h(x,0)=mx$ and the helicoidal boundary condition
$h(L+i,t)=h(i,t)+Lm$. Numerical results are
discussed below in detail. \\

\begin{figure}
\centerline{\epsfxsize=6cm \epsfbox{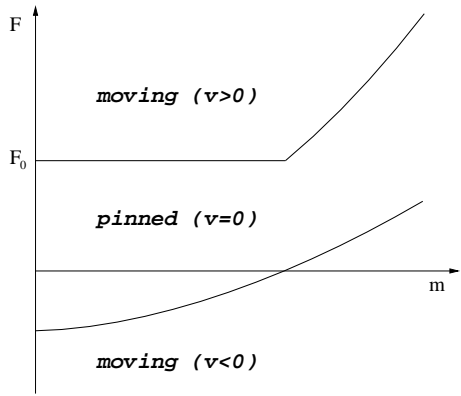}}
\caption{The phase diagram in the $F$-$m$ plane.}
\label{fig3}
\end{figure}

First, we measure the critical driving force $F_0$ and the facet slope at $m$=0
(no tilt) for several values of $\lambda$. $F_0$ is numerically defined as the
value of $F$ at which no surface in the sample of upto 200 configurations
 is pinned until a large fixed time $t_0 \sim 10^6 \Delta t  $. Just below $F_c$, 
at least a finite fraction of the ensemble 
is pinned at $t_0$ showing the facet morphology. 
The facet slope is 
determined by sample averages of 100 configurations.  
Measured values of $F_0$ and $s_c$ are shown in Table 1 for 
$\lambda= -0.5, -1, -1.5$ and $-2$.
 Also shown are $v_0$, the velocity at $F_0$, i.e. the velocity 
discontinuity at the first order PD transition.  
Next, we examine the tilt-dependence of the surface growth. For small tilts,
 the first order nature of the PD transition and the morphology of 
the pinned surfaces at the transition do not change.  
 Moreover, within our numerical accuracies, both 
the critical driving force and the critical facet slope are 
 independent of the tilt $m$.  We find this is so as long as $m$ is less than
a critical tilt $m_c$ which is essentially the same as the facet slope $s_c$.
When $m$ becomes larger than 
$s_c$,  the surfaces at the pinned state 
cannot form a facet with
slope $s_c$ by the helicoidal boundary condition and instead align 
along the substrate. In this case, the critical driving force $F_c(m)$ 
has to increase with $m$ to compensate for the tilt-force.
The schematic phase diagram in the $F-m$ plane is shown in Fig.\ref{fig3}. 
The positively moving phase is bounded by a horizontal line at $F=F_0$
for $m<m_c$ and a smooth curve $F_c(m)$ for $m>m_c$. 
The transition to the $v<0$ phase (the lower curve in Fig.\ref{fig3}) is the continuous
PD transition of directed percolation universality class.
This region is better understood by rewriting the QKPZ
equation in term of $h'\equiv -h$ as
\begin{equation}
{\partial_t h'}=\nu{\partial_x^2 h'}-{\lambda \over 2}(\partial_x h')^2
-F+\eta(x,h').
\end{equation}
The above equation implies that the negative QKPZ equation
with negative $F$ is equivalent to the positive QKPZ equation
with positive $F$ and vice versa.
The negative velocity phase in $F>0$ region of Fig.\ref{fig3}
can be understood as a result of the positive
QKPZ equation driven by a negative force which is not strong enough to overcome
the effect of the positively driving non-linear term. We confine our discussion
to the upper curve of Fig.\ref{fig3} below.  \\
 
\begin{figure}
\centerline{\epsfxsize=7.3cm \epsfbox{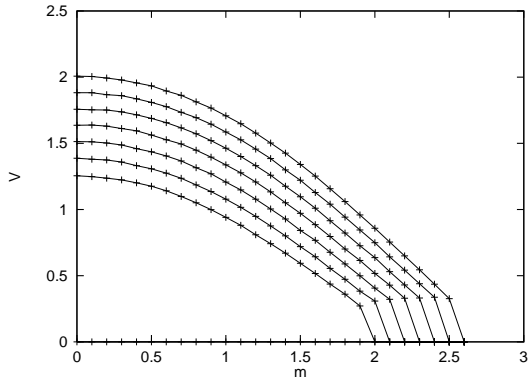}}
\vspace{.2cm}
\caption{The velocity versus the substrate-tilt $m$ for several 
external driving forces $F$ for the case of $\lambda=-0.5$. 
The curve at the bottom corresponds to  the magnitude of the velocity jump 
at the critical force $F_0$, which becomes zero at a 
characteristic slope $m_c$.} 
\label{fig4}
\end{figure}

\begin{figure}
\centerline{\epsfxsize=7.3cm \epsfbox{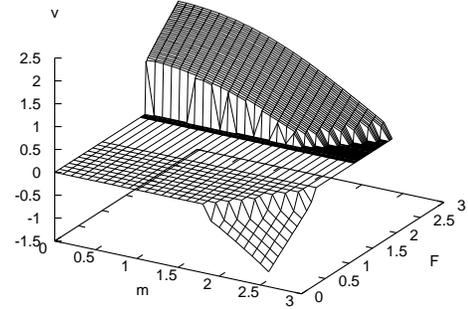}}
\vspace{.3cm}
\caption{Three dimensional plot of the velocity versus 
substrate-tilt $m$ and external force $F$ for the case of $\lambda=-0.5$.}
\label{fig5}
\end{figure}

\begin{figure}
\centerline{\epsfxsize=7.3cm \epsfbox{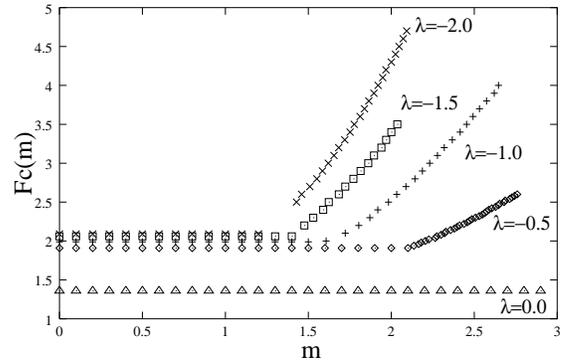}}
\vspace{.3cm}
\caption{The depinning critical $F_c(m)$ versus the substrate-tilt 
$m$ for several values of $\lambda$. }
\label{fig6}
\end{figure}

\begin{figure}
\centerline{\epsfxsize=7.3cm \epsfbox{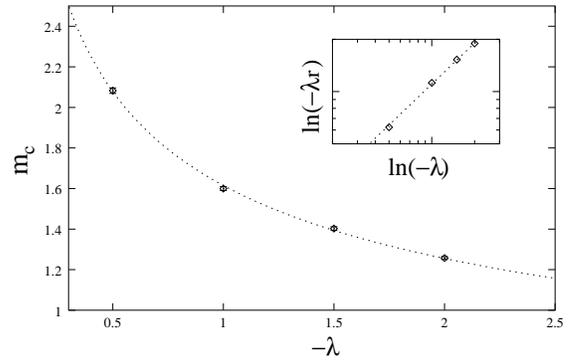}}
\vspace{.3cm}
\caption{Estimated numerical data for the critical tilt $m_c$ 
versus $\lambda$. The broken line $m_c \sim |\lambda|^{-0.37}$ 
is a least square fit. Inset: The $\lambda$-dependence of
$\lambda_r$ showing $|\lambda| \sim |\lambda_r|^{0.8}$.}
\label{fig7}
\end{figure}

A typical data set for the velocity as a function of $F$ and $m$ in 
the positively moving phase
is shown in Fig.\ref{fig4}, while Fig.\ref{fig5} shows a three dimensional 
plot of $v$ versus $F$ and $m$. At $F=F_0$ and for $m<m_c$,
$v$ exhibits a discontinuous jump by $\delta v(m)$. The lowest curve in 
Fig.\ref{fig4} shows this. Since  $\delta v(m)$ 
 vanishes at $m$=$m_c$, 
 the point ($F$=$F_0$,$m$=$m_c$) 
may be regarded as a tricritical point in analogy with the 
equilibrium critical phenomena.
Following \cite{abs},
we measure $\lambda_r$ by fitting the curves $v(F,m)$ to Eq.(\ref{vfm}) 
for small
$m$. Since it is insensitive to $F$, we define $\lambda_r$ as that obtained
from $F=F_0$;
\begin{equation}\label{vm}
\delta v(m) \equiv v(F_0,m) = v_0 - {|\lambda_r|\over 2} m^2 + \cdots .
\end{equation}
The fifth column of Table 1 shows $\lambda_r$.
For $F$ larger than 
but close to $F_0$, $v(F,0)$ is linear in $F$. 
This together with
Eq.(\ref{vm}) allows one to approximate $v(F,m)$ as 
\begin{equation} \label{vfm2}
v(F,m) = v_0 + A(F-F_0) - {|\lambda_r|\over 2} m^2 +\cdots
\end{equation}
where $A$ is a constant.
Setting $v=0$ in Eq.(\ref{vfm2}), one obtains an approximate form for
$F_c(m)$ as 
\begin{equation} \label{fcm}
F_c(m) \sim F_0 +{ |\lambda_r| \over {2A}} (m^2 - m_c^2) 
\end{equation} 
with $m_c = (2 v_0/|\lambda_r|)^{1/2}$.
 However, for larger $m$, $v$ deviates from 
Eq.(\ref{vfm2}) and higher order terms in $m$ become significant.
To find $F_c(m)$ and $m_c$ more accurately, 
we fit the velocity curve for large $m$ and
estimate $F_c(m)$ independently from the condition of $v(F_c(m),m)=0$.
This is justified since $\theta=1$ generically for $m \ne 0$ \cite{dhar}.
In Fig.\ref{fig6}, we show the values of $F_c(m)$ obtained in this way. The 
relation $F_c(m_c)=F_0$ then gives an independent estimate of $m_c$ which is
also shown in Table 1. Note that $m_c$ obtained in this way agrees well with 
the facet slope $s_c$ as mentioned above. 
The values of $m_c$ itself strongly depends on $\lambda$ and could be
fitted to a power law as 
$m_c \sim |\lambda|^{-0.37}$ as shown in Fig.\ref{fig7}. 
Thus when $\lambda=0$, $m_c$ becomes infinite, 
which recovers the previous result that the critical 
force $F_c(m)$ is independent of the substrate-tilt 
in the quenched Edwards-Wilkinson universality class \cite{dhar,abs}.
On the other hand, $|\lambda_r|$ scales as $|\lambda|^{0.8}$ in Fig.\ref{fig7} and hence 
$m_c \sim |\lambda|^{-0.37} \sim |\lambda_r|^{-0.46}$. This is to be compared with the zeroth order
expression $m_c \sim |\lambda_r|^{-0.5}$ in Eq.(\ref{fcm}). \\

In order to understand the origin of the discontinuous 
PD transition for $\lambda < 0$, we examine the noise distribution 
on the perimeter sites of critically pinned surfaces.  
It is found that the noise strengths around the site at the valley 
in Fig.\ref{fig2} are relatively small, 
corresponding to large pinning strengths. 
Meanwhile, when $\lambda <0$, surfaces tend to form a flat facet 
to make the term $(\lambda/2)(\partial_x h)^2$ negatively large, 
leading to a large pinning strength. 
However, the curvature force $\nabla^2 h$ 
is almost zero on the hillside, but is  positive 
at the valley. Thus the sites around the valley are pinned 
by balancing the driving forces due to the curvature and the external 
force $F$ with the pinning forces due to the noise and   
the non-linear term. Once the sites around the valley is pinned, 
the stable slope $s_c$ of facet is selected in a self-organized way: 
Suppose that a surface is formed with  a facet of slope $s < s_c$. 
Then the velocity of the sites on the hillside would be positive 
according to Fig.\ref{fig4} The pinning strength due to $(\lambda/2)s^2$ 
is not strong enough to resist the external force, and the sites 
on the hillside move upward, and tend to make the slope of the facet 
larger. When the slope becomes larger than $s_c$, however, 
the non-linear term is too strong, and the sites on the hillside move downward 
reducing the slope. Thus the slope $s_c$ becomes stable, and 
surfaces form a facet with the slope $s_c$ as shown in Fig.\ref{fig2}.    
Since the surface pinning is initiated at the sites around the 
valley, once the pinning site is broken by an external force which is slightly
bigger than $F_0$,
the surface grows with a finite velocity, leading to the first order transition.
On the other hand, when $\lambda > 0$, the non-linear term is positive, 
and enhances the external driving force. Thus surfaces do not 
form a facet. The surface pinning is mainly due to local noise 
strengths as can be pictured in the DP cluster. 
Thus the pinned sites are not localized but scattered all over the system,
so that the PD transition is continuous.\\    
 
It would be interesting to observe the tilt-dependence 
of the negative quenched KPZ equation from a stochastic model. 
The Sneppen A model \cite{noise,sneppen} is known to belong 
to the negative QKPZ universality class. 
In the Sneppen A model, the restricted solid-on-solid (RSOS) condition 
is imposed on the height difference of neighboring columns.    
Because of this restriction, the surface tilt cannot be larger than a
maximum value. 
If the maximum tilt is less than the critical slope $s_c(\lambda)$ 
of the stochastic model, the tricritical behavior 
would not be observed in simulations 
because the coefficient $\lambda$ is not controllable. 
Recently a PD transition similar to 
the negative QKPZ equation is found in a seemingly different system, 
the driven Frenkel-Kontorova model 
\cite{fk}, where the PD transition is also discontinuous and 
the displacement at the pinned state looks like Fig.\ref{fig2}. 
However, the tricritical behavior has not been studied yet 
in this system.\\ 

In summary, we have investigated the tilt-dependent behavior of 
the negative quenched QKPZ equation. We observed and explained 
that there exists a characteristic substrate-tilt $m_c$ such that
the PD transition is discontinuous
(continuous) when the substrate-tilt $m$ is less (greater) than $m_c$.
 The characteristic tilt $m_c$ is found to be the same as the facet
slope $s_c$ of the critically pinned surfaces which are formed in 
self-organized way and depends on the non-linear term coefficient $\lambda$ as 
$m_c \sim |\lambda|^{-0.37}$. 
Moreover, the threshold force $F_c$ for the PD transition is 
independent of $m$ for $m < m_c$ and increases with increasing 
$m$ for $m > m_c$. The effective non-linear term coefficient $\lambda_r$ 
remains finite as $F$ approaches $F_c$.\\    

We would like to thank M. Kardar for suggesting this problem 
and F.-J. Elmer 
for sending Ref.\cite{fk}. This work was supported in part by the KOSEF 
through the SRC program of SNU-CTP, and in part by the Ministry 
of Education, Korea (97-2409). \\ 
                                                  

\vspace{-1.5cm}

\begin{table}
\caption{Measured values of the critical driving force $F_0$,
the facet slope $s_c$, the velocity discontinuity $v_0$,
the effective non-linear parameter $\lambda_r$, and the critical
substrate-tilt $m_c$. }
\begin{center}
\begin{tabular}{cccccc}
  $\lambda$ & $F_0$ & $s_c$ & $v_0$ &$\lambda_r$& $m_c$ \\
\hline
  -0.5 & 1.91 & 2.08 & 1.20 & -0.62 & 2.11\\
  -1.0 & 1.99 & 1.60 & 1.30 & -1.12 & 1.64\\
  -1.5 & 2.06 & 1.40 & 1.35 & -1.52 & 1.39\\
  -2.0 & 2.09 & 1.26 & 1.42 & -1.89 & 1.24 
\end{tabular}
\end{center}
\end{table}

\end{multicols}
\end{document}